\documentclass[12pt, fleqn]{article}
\usepackage[cp1251]{inputenc}
\usepackage{latexsym,amsfonts,amssymb}
\usepackage{graphicx}

\usepackage{amsbsy}
\usepackage{amsmath}
\usepackage{epsf}
\usepackage{cite}

\newtheorem{theo}{Theorem}
\newcommand{\bt}{\begin{theo}}
\newcommand{\et}{\end{theo}}
\newcommand{\bd}{\begin{displaymath}}
\newcommand{\ed}{\end{displaymath}}

\newcommand{\lf}{\left}
\newcommand{\rg}{\right}

\newcommand{\be} {\begin{equation}}
\newcommand{\ee} {\end{equation}}
\newcommand{\ba}{\begin{array}{l}}
\newcommand{\ea} {\end{array}}

\newcommand{\p} {\partial}

\begin{document}
 \begin{center}
 {\Large \bf
 Nonlinear   reaction-diffusion systems with a non-constant \\ diffusivity: conditional symmetries in no-go case   }\\
 \medskip

{\bf Roman Cherniha$^{\dag,\ddag}$\footnote{\small  Corresponding author. E-mails: cherniha@gmail.com; Roman.Cherniha@nottingham.ac.uk}}
 {\bf and  Vasyl' Davydovych$^\dag$}
 \\
{\it  $^\dag$~Institute of Mathematics, Ukrainian National Academy
of Sciences,\\
 3 Tereshchenkivs'ka Street, Kyiv 01601, Ukraine\\
  $^\ddag$~\it School of Mathematical Sciences,
University of Nottingham, \\
University Park, Nottingham, NG7 2RD, UK
 }\\
\end{center}

\begin{abstract}
 $Q$-conditional symmetries (nonclassical symmetries)
 for a general  class of two-component reaction-diffusion   systems  with non-constant diffusivities  are studied.
  The work is a natural continuation of our paper ``Conditional symmetries and exact solutions of nonlinear
reaction–diffusion systems with non-constant
diffusivities''(Cherniha and Davydovych, 2012) \cite{che-dav2012} in
order to extend the  results on so-called no-go case.
 Using the    notion of  $Q$-conditional symmetries
 of the first type, an  exhaustive list of  reaction-diffusion  systems  admitting  such symmetry is derived. The results obtained   are compared with those derived earlier. The  symmetries     for  reducing  reaction-diffusion   systems to
  two-dimensional dynamical systems (ODE systems)  and finding exact solutions are  applied.
 As result,  multiparameter  families  of exact solutions in the explicit  form
 for  nonlinear   reaction-diffusion  systems   with an arbitrary power-law diffusivity are  constructed  and their properties for possible applicability are established.
\end{abstract}

\textbf{Keywords:}  nonlinear reaction-diffusion system; Lie symmetry;  non-classical symmetry;
$Q$-conditional symmetry  of the first type;   exact solution.

\newpage

\section{\bf Introduction}

This work is a natural continuation of our paper \cite{che-dav2012} and
 is devoted  to the  investigation of
 the two-component reaction-diffusion (RD)  systems of the form
 \be\label{1}\ba
 U_t=[D^1(U)U_x]_x+F(U,V),\\
V_t=[D^2(V)V_x]_x+ G(U,V),
 \ea\ee
where
  $U= U(t,x)$ and $V= V(t,x)$ are two  unknown functions representing, say,  the  densities
  of populations (cells, chemicals),
$F(U,V)$ and $G(U,V)$ are the  given smooth functions describing interaction
between them and environment,
 the functions $D^1(U)$ and $D^2(V)$ are the relevant diffusivities
 (hereafter they are  positive smooth  functions)  and the subscripts $t$ and $x$ denote
differentiation with respect to these variables.
 The class of  RD systems  (\ref{1}) generalizes many well-known nonlinear
second-order models and is used to describe various processes in
physics, biology, chemistry  and ecology. The relevant examples can be found in  the well-known books
 \cite{ames, mur2, mur2003, okubo} and a wide range of papers.

 In paper \cite{che-dav2012},   the $Q$-conditional  invariance of these systems  in the case  when the operator in question has the form
  \be\nonumber Q = \xi^0 (t,
x, u, v)\p_{t} + \xi^1 (t, x, u, v)\p_{x} +
 \eta^1(t, x, u, v)\p_{u}+\eta^2(t, x, u, v)\p_{v},  \ee
 where $\xi^0\neq0$, has been examined and   an  exhaustive list of  reaction-diffusion  systems  admitting    $Q$-conditional symmetries
 of the first type \cite{ch-2010}  has been derived.  Here we study so-called no-go case when $\xi^0 =0$, which is thought
 to be much more difficult  and usually is skipped from examination.
  The additional  reason to avoid  examination of no-go case was the well-known  fact
  (firstly    proved in  \cite{zh-lahno98}) that complete description of  $Q$-conditional symmetries with $\xi^0 =0$ for scalar evolution equations is
  equivalent to  solving of the equation in question.

On the other hand,  the algorithm  of heir-equations was
 proposed in \cite{nuc-94}, which
 allows  to construct a hierarchy of the conditional symmetry operators  starting from a particular one with $\xi^0 =0$.
 This algorithm was successfully applied in order to find exact solutions for some evolution equations
  (in particular, see its application in  the  recent paper \cite{hashemi-nuc-13}).
  It can be  also noted the very recent paper \cite{Ji-Qu-Shen-14}, in which  RD systems
   were investigated in order to find conditional Lie-B\"acklund symmetry  (generalised
    conditional symmetries in terminology of the pioneering paper \cite{fok-liu-94}). However, this is well-known that deriving a complete
classification of generalised conditional symmetries is unrealistic
even for classes of scalar PDEs because the relevant systems of
determining equations are very complicated.

 To the best of our knowledge, there are no papers devoted to search
 $Q$-conditional symmetry (non-classical symmetry) of the  class of  systems
  (\ref{1}) in the case $\xi^0 =0$ and application of such symmetries for finding exact solutions of nonlinear reaction-diffusion  systems.
  Here we show that a complete description of  $Q$-conditional symmetries
 of the first type \cite{ch-2010} can be  derived in no-go case.

 The paper is organized as follows. In  section~2,
  basic   definitions
   are presented, the systems of   determining
   equations are derived  and the main theorems are proved.
  In section~3, the $Q$-conditional  symmetries obtained for  reducing of
   RD systems to systems of  ODEs  are applied  and  multiparameter  families
    of exact solutions are constructed. Moreover, it is shown that the solutions obtained
     possess attractive properties, which may lead to their possible applications.
    Finally, we  summarize and discuss
    the results obtained   in
the last section.

\section{\bf Conditional symmetry for RD systems }

\subsection{Definitions and preliminary analysis}

Following  \cite{che-dav2012}, we simplify RD system  (\ref{1})  by applying  the Kirchhoff
substitution \be\label{2} u = \int D^1(U)dU, \quad v = \int
D^1(V)dV, \ee where $u(t,x)$ and $v(t,x)$ are new unknown functions.
Hereafter we assume that there exist unique  inverse functions to those
arising in right-hand-sides of (\ref{2}). Substituting (\ref{2})
into (\ref{1}), one obtains
\be\label{b2}\ba
u_{xx}=d^1(u)u_t+C^1(u,v),
\\v_{xx}=d^2(v)v_t+C^2(u,v),\ea\ee
where the functions  $d^1,\ d^2, \ C^1$ and $C^2$ are uniquely
defined via $D^1,\ D^2, \ F$ and $G$ by the formulae
\be\label{3*}d^1(u)=\frac{1}{D^1(U)}, \ d^2(v)=\frac{1}{D^2(V)}, \
C^1(u,v)=-F(U,V), \ C^2(u,v)=-G(U,V),\ee where  $U =D^1_*(u) \equiv
\lf(\int D^1(u)du \rg)^{-1} , \quad V=D^2_*(v) \equiv \lf(\int
D^2(v)dv \rg)^{-1}$ (the upper subscripts $-1$ mean  inverse
functions).

Hereafter we  examine  class of RD systems   (\ref{b2}) instead of
(\ref{1}) because   both classes are  equivalent. In fact,   having
any conditional symmetry operator of a RD system of the form
(\ref{b2}), one may easily transform those  into the relevant
operator and a  RD system from the class  (\ref{1}) provided the
inverse functions in  (\ref{3*}) are known.

It is well-known that to find Lie invariance  operators, one needs
to consider   system (\ref{b2}) as the manifold
${\cal{M}}=\lf\{S_1=0,S_2=0 \rg\}$  where \be\nonumber  \ba
 S_1 \equiv \ u_{xx}-d^1(u)u_t-C^1(u,v),\\
S_2 \equiv \ v_{xx}-d^2(v)v_t-C^2(u,v), \ea\ee in the prolonged
space of the  variables: $t, x, u, v, u_t, v_t$,$ u_{x}, v_{x},
u_{xx}, v_{xx}, u_{xt}, v_{xt}, u_{tt}, v_{tt}.$ According to the
definition, system (\ref{b2}) is invariant under the transformations
generated by the infinitesimal operator
 \be\label{b3} Q = \xi^0 (t,
x, u, v)\p_{t} + \xi^1 (t, x, u, v)\p_{x} +
 \eta^1(t, x, u, v)\p_{u}+\eta^2(t, x, u, v)\p_{v},  \ee
if the following invariance conditions are satisfied: \be\nonumber
\ba \mbox{\raisebox{-1.5ex}{$\stackrel{\displaystyle Q}{\scriptstyle
2}$}}\, \lf(S_1\rg)\Big\vert_{\cal{M}}
 =0, \\[0.3cm]
\mbox{\raisebox{-1.5ex}{$\stackrel{\displaystyle Q}{\scriptstyle
2}$}}\, \lf(S_2\rg)\Big\vert_{\cal{M}}
 =0. \ea \ee
The operator $ \mbox{\raisebox{-1.5ex}{$\stackrel{\displaystyle  
Q}{\scriptstyle 2}$}} $  
is the second  
 prolongation of the operator $Q$, i.e.  
\be\nonumber\ba  
\mbox{\raisebox{-1.5ex}{$\stackrel{\displaystyle  
Q}{\scriptstyle 2}$}}  
 = Q + \rho_t^1\frac{\partial}{\partial u_{t}}+
\rho^1_x\frac{\partial}{\partial
u_{x}}+\rho_t^2\frac{\partial}{\partial v_{t}}+
\rho^2_x\frac{\partial}{\partial v_{x}}
+\\\hskip1cm\sigma_{tt}^1\frac{\partial}{\partial
u_{tt}}+\sigma_{tx}^1\frac{\partial}{\partial u_{tx}}+
\sigma_{xx}^1\frac{\partial}{\partial u_{xx}}
+\sigma_{tt}^2\frac{\partial}{\partial
v_{tt}}+\sigma_{tx}^2\frac{\partial}{\partial  v_{tx}}+
\sigma_{xx}^2\frac{\partial}{\partial  v_{xx}},  
\ea\ee  
where the coefficients $\rho$ and $\sigma$ with relevant subscripts  
are expressed  via the functions $\xi^0, \xi^1, \eta^1$ and $\eta^2$
by well-known formulae (see, e.g., \cite{fss, olv, b-k}).

The crucial idea used for introducing the notion of $Q$-conditional
symmetry (non-classical symmetry) is to change  the manifold
${\cal{M}}$, namely: the operator $Q$ is used to reduce ${\cal{M}}$ (see the pioneer  paper  \cite{bl-c}).
However,
 there are two essentially different
possibilities to realize this idea in the case of two-component
systems. Moreover, there are many different possibilities in the
case of multi-component systems  \cite{ch-2010}. Following
\cite{ch-2010}, we formulate two definitions  in the case of system
(\ref{b2}).

\noindent \textbf{ Definition 1.} Operator (\ref{b3}) is called the
$Q$-conditional symmetry of the first type  for the RD system
(\ref{b2}) if  the following invariance conditions are satisfied:
\be\nonumber \ba \mbox{\raisebox{-1.5ex}{$\stackrel{\displaystyle
Q}{\scriptstyle 2}$}}\, \lf(S_1\rg)\Big\vert_{{\cal{M}}_1}
 =0, \\[0.3cm]
\mbox{\raisebox{-1.5ex}{$\stackrel{\displaystyle Q}{\scriptstyle
2}$}}\, \lf(S_2\rg)\Big\vert_{{\cal{M}}_1}
 =0, \ea \ee where the manifold ${\cal{M}}_1$
is either $\left\{S_1=0, S_2=0, Q(u)=0 \right\}$ or $\{S_1=0, S_2=0,
Q(v)=0 \}$.

\noindent \textbf{Definition 2.} Operator (\ref{b3}) is called the
$Q$-conditional symmetry of the second  type, i.e., the standard
non-classical symmetry  for the RD system (\ref{b2}) if  the
following invariance conditions are satisfied: \be\nonumber \ba
\mbox{\raisebox{-1.5ex}{$\stackrel{\displaystyle Q}{\scriptstyle
2}$}}\, \lf(S_1\rg)\Big\vert_{{\cal{M}}_2}
=0, \\[0.3cm]
\mbox{\raisebox{-1.5ex}{$\stackrel{\displaystyle Q}{\scriptstyle
2}$}}\, \lf(S_2\rg)\Big\vert_{{\cal{M}}_2} =0, \ea \ee where the
manifold ${\cal{M}}_2=\{S_1=0, S_2=0, Q(u)=0, Q(v)=0 \}$.

It is well-known that constructing $Q$-conditional symmetries for evolution equations and evolution systems leads to the requirement
to analyze two essentially different cases
 {\it a)} $\xi^0\neq0;$ \  {\it b)} $\xi^0=0, \
\xi^1\neq0$  (see operator (\ref{b3})).  The case {\it a)} for   system (\ref{b2}) was examined in our previous  papers  \cite{ch-pli-08, che-dav2012, che-dav2014}. Here we restrict ourselves on the case {\it b)}, i.e. when operator (\ref{b3}) has the structure
\be\label{b1} Q = \xi(t, x, u, v)\p_{x} +
 \eta^1(t, x, u, v)\p_{u}+\eta^2(t, x, u, v)\p_{v}.\ee

 In order to find $Q$-conditional symmetries of the second
  type (non-classical symmetries),  one applies  Definition~2
  to system (\ref{b2}). Direct calculations  produce the following system of determining equations:
\begin{eqnarray}&&
\nonumber
\left(d^1-d^2\right)\left(\xi\eta^1_{v}-\eta^1\xi_v\right)=0,\quad
 \left(d^1-d^2\right)\left(\xi\eta^2_{u}-\eta^2\xi_u\right)=0,\\
&&\nonumber
2d^1\left(\xi_x+\frac{\eta^1}{\xi}\xi_u+\frac{\eta^2}{\xi}\xi_v\right)
 +\eta^1d^1_u =0,\\
&&\nonumber
2d^2\left(\xi_x+\frac{\eta^1}{\xi}\xi_u+\frac{\eta^2}{\xi}\xi_v\right)
 +\eta^2d^2_v =0,\\
&&\nonumber\eta^1C^1_u+\eta^2C^1_v+\left(\frac{\eta^1}{\xi}
\xi_v-\eta^1_v\right)C^2+\left(2\xi_x
+3\frac{\eta^1}{\xi}\xi_u+2\frac{\eta^2}{\xi}\xi_v-
\eta^1_u\right)C^1
+\left(\eta^1_t-\frac{\eta^1}{\xi}\xi_t\right)d^1-\\
&&\label{b57}
\eta^1_{xx}+\frac{\eta^1}{\xi}\left(\xi_{xx}-2\eta^1_{xu}\right)+
\left(\frac{\eta^1}{\xi}\right)^2\left(2\xi_{xu}-\eta^1_{uu}\right)+
2\frac{\eta^1\eta^2}{\xi^2}\left(\xi_{xv}- \eta^1_{uv}\right)+
\\ && \nonumber\left(\frac{\eta^1}{\xi}\right)^3\xi_{uu}+
\frac{\eta^1(\eta^2)^2}{\xi^3}\xi_{vv}+
2\frac{\eta^2(\eta^1)^2}{\xi^3}\xi_{uv}-2\frac{\eta^2}{\xi}\eta^1_{xv}-
\left(\frac{\eta^2}{\xi}\right)^2\eta^1_{vv}=0 ,\\
&&\nonumber\eta^1C^2_u\eta^2C^2_v+\left(\frac{\eta^2}{\xi}
\xi_u-\eta^2_u\right)C^1+\left(2\xi_x
+3\frac{\eta^2}{\xi}\xi_v+2\frac{\eta^1}{\xi}\xi_u-
\eta^2_v\right)C^2
+\\
&&\nonumber\left(\eta^2_t-\frac{\eta^2}{\xi}\xi_t\right)d^2-
\eta^2_{xx}+\frac{\eta^2}{\xi}\left(\xi_{xx}-2\eta^2_{xv}\right)+
\left(\frac{\eta^2}{\xi}\right)^2\left(2\xi_{xv}-\eta^2_{vv}\right)+
2\frac{\eta^1\eta^2}{\xi^2}\left(\xi_{xu}- \eta^2_{uv}\right)+
\\ && \nonumber\left(\frac{\eta^2}{\xi}\right)^3\xi_{vv}+
\frac{\eta^2(\eta^1)^2}{\xi^3}\xi_{uu}+
2\frac{\eta^1(\eta^2)^2}{\xi^3}\xi_{uv}-
2\frac{\eta^1}{\xi}\eta^2_{xu}-
\left(\frac{\eta^1}{\xi}\right)^2\eta^2_{uu}=0.
\end{eqnarray}
Hereafter the subscripts $t$, $x$, $u$ and $v$ denote
differentiation with respect to these variables.
\medskip

  \textbf{Remark 1.} One cannot set  $\xi=1$  in operator (\ref{b1}) without losing a generality  for
   obtaining  the system of determining equations (in contrary to the case of operator (\ref{b3})
    with  $\xi^0\neq0$!)
  because the system obtained immediately leads to the trivial result $Q = \p_{x} $.

One easily notes that system (\ref{b57}) consists of 6 nonlinear PDEs for 7 unknown functions  $d^1(u), \ d^2(v), \ C^1(u,v), \ C^2(u,v), \
\xi(t, x, u, v), \
 \eta^1(t, x, u, v)$  and  $\eta^2(t, x, u, v)$. In order to solve any system of determining equations one needs to establish structure of coefficients of operator (\ref{b1}). In the case of system (\ref{b57}),
 it means that the subsystem consisting of four PDEs for  $d^1(u), \ d^2(v), \  \xi(t, x, u, v), \
 \eta^1(t, x, u, v)$  and  $\eta^2(t, x, u, v)$ should be examined. Unfortunately,
 we were unable to solve this subsystem for arbitrary functions  $d^1(u)$ and $ d^2(v) $
 and believe that it is possible to do for  the correctly-specified pairs $(d^1(u),d^2(v))$ only.

Happily  our efforts were successful in application of   Definition
1. Formally speaking, we should construct  systems of DEs  using two
different manifolds  ${\cal{M}}_1$ (see Definition~1). However, the
class of RD systems  (\ref{b2})
 is invariant under discrete
transformation $u \to v, \, v \to u$. Thus, we can use only one
manifold, say,
 $\lf\{S_1=0, S_2=0, Q(u)=0 \rg\}$. Having the complete list of the conditional
 symmetry operators and the relevant forms of RD systems, one may simply
 extend such list by applying the transformation mentioned above.

 Thus, now we present the system of DEs, obtained by direct application of  Definition~1 with
  ${\cal{M}}_1=\{S_1=0, S_2=0, Q(u)=0 \}$, for finding $Q$-conditional symmetry
  operator of the form (\ref{b1}), namely:
\begin{eqnarray}&&
\label{b4} \xi_{u}=\xi_{v}=0,\\
&& \label{b5} (d^1-d^2)\eta^1_{v}=(d^1-d^2)\eta^2_{u}=\eta^1_{vv}=
\eta^2_{vv}=0, \quad \eta^1_{xv}+\eta^1_{uv}\frac{\eta^1}{\xi}=0,\\
&&\label{b6} 2\xi_xd^1 +\eta^1d^1_u =0,\\
&&\label{b7} 2\xi_xd^2 +\eta^2d^2_v =0,\\
&&\label{b8} \xi_td^2+2\eta^2_{xv}-\xi_{xx}+2\frac{\eta^1}{\xi}\eta^2_{uv}=0,\\
&&\nonumber\eta^1C^1_u+\eta^2C^1_v-\eta^1_vC^2+(2\xi_x -\eta^1_u)C^1
+\eta^1_td^1-\eta^1_{xx}-\\ &&\label{b9}
\hskip2cm\left(\frac{\eta^1}{\xi}\right)^2\eta^1_{uu}-\frac{\eta^1}{\xi}\left(\xi_td^1+2\eta^1_{xu}
-\xi_{xx}\right)=0,\\
&&\nonumber
\eta^1C^2_u+\eta^2C^2_v-\eta^2_uC^1+(2\xi_x-\eta^2_v)C^2+\eta^2_td^2
 -\eta^2_{xx}-\\&& \label{b10}
 \hskip2cm \left(\frac{\eta^1}{\xi}\right)^2\eta^2_{uu}-2\frac{\eta^1}{\xi}\eta^2_{xu} =0.
\end{eqnarray}
It should be stressed that we are looking for  purely conditional symmetry
operators, i.e., all the  operators, which are equivalent to
the Lie symmetries  presented in
  \cite{ch-king4} should be excluded.  Having  this aim,  we use
  the DEs system to search for Lie symmetry operators of
  the form (\ref{b1}):
\begin{eqnarray}
& &\xi_{u}=\xi_{v}=0, \nonumber \\
& & \eta^1_{uu}=\eta^1_{uv}=\eta^1_{vv}=\eta^2_{uu}=\eta^2_{uv}=
\eta^2_{vv}=0, \nonumber \\
& & (d^1-d^2)\eta^1_{v}=0, \quad (d^1-d^2)\eta^2_{u}=0, \quad
\eta^1_{xv}=\eta^2_{xu}=0, \nonumber \\
& &2\xi_xd^1 +\eta^1d^1_u =0, \nonumber \\
& &2\xi_xd^2 +\eta^2d^2_v =0, \label{b11} \\
& &\xi_td^1+2\eta^1_{xu}-\xi_{xx}=0, \nonumber \\
& &\xi_td^2+2\eta^2_{xv}-\xi_{xx}=0, \nonumber \\
& &
\eta^1C^1_u+\eta^2C^1_v-\eta^1_vC^2+(2\xi_x-\eta^1_u)C^1+\eta^1_td^1
-\eta^1_{xx}=0,\nonumber \\
 & &\eta^1C^2_u+\eta^2C^2_v-\eta^2_uC^1+(2\xi_x-\eta^2_v)C^2+\eta^2_td^2
 -\eta^2_{xx}=0,\nonumber
\end{eqnarray} which can be easily extracted
from the relevant  system derived in    \cite{ch-king4}. One notes, that
systems of DEs
  (\ref{b4})--(\ref{b10})   and  (\ref{b11}) coincide (in the case $d^1\neq d^2$) if the restrictions
 \be\label{b12}
 \eta^1_{uu}=0, \quad
\xi_td^1+2\eta^1_{xu}-\xi_{xx}=0 \ee take place. Thus, we take into
account only such solutions of  (\ref{b4})--(\ref{b10}), which do not
satisfy at least one of the equations from  (\ref{b12}). Moreover, since
$Q$-conditional symmetry of the first type is automatically one of
the  second type, we should also check  the same for coefficients of
the operator obtained by multiplying  (\ref{b3}) on any  smooth
functions.  Otherwise the  $Q$-conditional symmetry obtained will be
equivalent to a Lie symmetry.

\subsection{The main theorems}

It turns out that to solve the DEs system (\ref{b4})--(\ref{b10}) one needs to examine  separately four cases
 \begin{enumerate}
            \item  two nonconstant diffusivities  $ \ d^1_ud^2_v\neq0$;
            \item  one nonconstant diffusivity  $  \ d^1_ud^2_v=0, \ (d^1_u)^2+(d^2_v)^2\neq0$;
             \item   two constant diffusivities being  different  $d^k=\lambda_k=const \ (k=1,2), \
\lambda_1\neq\lambda_2;$
              \item   two equal  diffusivities $d^k=\lambda_k=const \
(k=1,2), \ \lambda_1=\lambda_2$.
        \end{enumerate}

In fact, a simple analysis of  equations  (\ref{b5})--(\ref{b7}) shows that
 the form and  number of equations are  different   in each of the cases listed above.
 Similarly to the examination  of  operator (\ref{b3}) with $\xi^0 (t,
x, u, v) \not=0$ \cite{che-dav2012},  we consider firstly  two most  general cases,
involving at least one nonconstant diffusivity while other two cases will be treated  elsewhere.

Our main result can be
formulated in form of   two  theorems presenting  the complete lists of $Q$-conditional operators of the first type
and having the form (\ref{b1}), which are admitted by any RD system (\ref{b2}) with a nonconstant diffusivity.

\bt\label{bt1} In the case $d^1_ud^2_v\neq0$ the RD system
(\ref{b2}) admits  $Q$-conditional operator of the first type
(\ref{b1}), up to equivalent transformation \be\label{b35}\ba t\rightarrow C_1t+C_2,\quad x\rightarrow C_3x+C_4, \\
u\rightarrow C_5u+C_6,\quad v\rightarrow C_7v+C_8, \ea\ee (here $C_l
\ (l=1,\dots,8)$ are constants, $C_{2k-1}\neq0, \ k=1,\dots,4$) and
to discrete transformation \be\label{b36} u\rightarrow v, \quad
v\rightarrow u,\ee only in three cases: \be\label{b30}(I) \quad\ba
u_{xx}=d^1(u)\,u_t+\frac{\lf(d^1\rg)^2}{d^1_u}f\left(v^4d^1\right)+16\left(1-\frac{d^1d^1{uu}}{(d^1_u)^2}\right)\frac{d^1}{d^1_u},
\\v_{xx}=v^{-4}v_t+v^{-3}g\left(v^4d^1\right)+v,\ea\ee
\be\label{b31} Q = e^{2x}\left(\p_{x}-
 4\frac{d^1}{d^1_u}\,\p_{u}+v\p_{v}\right),\ee
where $d^1\neq \delta_1 u^{-4}$ is an arbitrary function;
\be\label{b32*}(II) \quad\ba
u_{xx}=d^1(u)\,u_t+\frac{\lf(d^1\rg)^2}{d^1_u}\left(f\left(v^4d^1\right)+\frac{4\mu}{d^1}
+4\mu\int\frac{d^1{uu}}{d^1d^1_u}du\right),
\\v_{xx}= v^{-4}v_t+v^{-3}g\left(v^4d^1\right)+\frac{\mu}{4}\,v,\ea\ee
\be\label{b33*} Q = \xi(x)\p_{x}-
 2\xi_x\frac{d^1}{d^1_u}\,\p_{u}+\frac{1}{2}\,\xi_xv\p_{v}, \ee where
  $d^1=e^u$ or $d^1=u^\beta \ (\beta\neq-4)$,
  while
\be\label{b49*}\xi(x)=
\begin{cases} \exp\left(\sqrt{\mu}\,x\right)+\alpha
\exp\left(-\sqrt{\mu}\,x\right),
  & $if$ \quad  \mu>0,
 \\ \sin\sqrt{-\mu}\,x  ,& $if$
  \quad \mu<0,
\end{cases} \ee where $\alpha\neq0$ is an arbitrary constant;
 \be\nonumber(III) \quad\ba
u_{xx}=d^1(u)\,u_t+\frac{\lf(d^1\rg)^2}{d^1_u}\left(f\left(v^4d^1\right)+\frac{3\mu}{2d^1}
+2\mu\int\frac{d^1{uu}}{d^1d^1_u}du\right),
\\v_{xx}= v^{-4}v_t+v^{-3}g\left(v^4d^1\right)+\frac{\mu}{4}\,v,\ea\ee
\be\nonumber Q = \xi(x)\p_{x}-
 2\xi_x\frac{d^1}{d^1_u}\,\p_{u}+\frac{\xi_x}{2}\,v\p_{v}, \ee where
  $d^1\neq\delta_1u^{-4}$ is an  arbitrary
  solution of the equation \be\nonumber
8\frac{d^1}{d^1_u}\left(\frac{d^1}{d^1_u}\right)_{uu}-
4\left(\frac{d^1}{d^1_u}\right)_u-1=0, \ee while
 \be\label{b49**}\xi(x)=
\begin{cases} x^2,
  &  $if$ \quad \mu=0,\\
 \left(\exp\left(\frac{\sqrt{\mu}}{2}\,x\right)+\alpha
\exp\left(-\frac{\sqrt{\mu}}{2}\,x\right)\right)^2, & $if$ \quad
 \mu>0, \\
 \sin\sqrt{-\mu}\,x\pm1, & $if$ \quad
 \mu<0,
 \end{cases} \ee where $\alpha\neq0$ is an arbitrary constant. In cases (I)--(III), $f$ and $g$ are arbitrary smooth functions of the argument $v^4d^1(u)$, while the function $d^1(u)$ is
  described in each case.
\et

\bt\label{bt2} In the case $d^1_ud^2_v=0$ and
$(d^1_u)^2+(d^2_v)^2\neq0$ the RD system (\ref{b2}) admits
$Q$-conditional operator of the first type (\ref{b1}), up to
equivalent transformation (\ref{b35}) and to discrete transformation
(\ref{b36}), only in such cases: \be\nonumber(I) \hskip1cm\ba
u_{xx}=d^1(u)u_t+f(u),\\v_{xx}=v_t+vg(u)+\alpha v\ln v,\ea\ee
\be\nonumber Q = e^{-\alpha t}(2\p_{x}+\alpha xv\p_{v}), ; \ee
\be\nonumber(II) \hskip8.1mm\ba
u_{xx}=d^1(u)u_t+f(u),\\v_{xx}=v_t+vg(u),\ea\ee \be\nonumber Q =
-2t\p_{x}+xv\p_{v};  \ee \be\nonumber(III) \hskip6.1mm\ba
u_{xx}=d^1(u)u_t+f(u),\\ v_{xx}=v_t+e^vg(u)+\alpha e^{2v},\ea\ee
\be\nonumber Q = \p_{x}+\alpha e^u\p_{u},  \ee
where $ \alpha\neq0 $.
In cases (I)--(III), $f$ and $g$ are arbitrary smooth functions of the argument $u$.
  \et

\textbf{Proofs  of Theorems 1  and 2}  is based on solving  the nonlinear DEs system
(\ref{b4})--(\ref{b10})  under restrictions $ \ d^1_ud^2_v\neq0$  and $  \ d^1_ud^2_v=0, \ (d^1_u)^2+(d^2_v)^2\neq0$,
respectively. We consider in details only  the proof of Theorem 1, which is more complicated.

 Differentiating equation (\ref{b7}) w.r.t.  $v$,  one immediately obtains $\xi_t=0, $  i.e. the
  function $\xi$ depends  only on $x$ (see Eqs.   (\ref{b4})).
 Now we solve equation  (\ref{b6})  and  (\ref{b7}) :
\begin{eqnarray}&&\label{b13}
\eta^1=-2\xi_x\frac{d^1}{d^1_u},\\&&\label{b14}
\eta^2=-2\xi_x\frac{d^2}{d^2_v}. \end{eqnarray}
and,  taking into account  restrictions  (\ref{b12}),
  conclude  that  $\xi_x\neq0.$

Because  $\eta^2_{vv}=0$  (see  (\ref{b5})) we  obtain from   (\ref{b14})  the first-order ODE
\be\label{b15}\frac{d^2}{d^2_v}=\alpha_1v+\alpha_2,\ee (here  $\alpha_1$
and $\alpha_2$   are arbitrary constants and $\alpha^2_1+\alpha^2_2\neq0$ )  with the general solutions

\begin{eqnarray}&&\label{b16}d^2=\delta_2\exp\left(\frac{v}{\alpha_2}\right),
\ \texttt{if} \ \alpha_1=0, \  \ \alpha_2\neq0,\\&&
\label{b17}d^2=\delta_2\left(\alpha_1v+\alpha_2\right)^{\frac{1}{\alpha_1}},
 \ \texttt{if} \ \alpha_1\neq0,\end{eqnarray} where  $\delta_2 \not=0$  is an arbitrary constant.

 A further examination of the remaining equations (\ref{b9})--(\ref{b10}) leads only to
 Lie symmetries provided the diffusivity $d^2$ has the form (\ref{b16}). Thus, we present
 only details for  $d^2$ of  the form (\ref{b17}).

Substituting  (\ref{b17})  and   (\ref{b13})--
(\ref{b15})  into the system in question and operator (\ref{b1}), one arrives at the system
\be\label{b18}\ba u_{xx}=d^1(u)u_t+C^1(u,v),
\\v_{xx}=\delta_2\left(\alpha_1v+\alpha_2\right)^{\frac{1}{\alpha_1}}v_t+C^2(u,v)\ea\ee
and the operator  \be\label{b19} Q = \xi(x)\p_{x}-
 2\xi_x\frac{d^1}{d^1_u}\,\p_{u}-2\xi_x(\alpha_1v+\alpha_2)\,\p_{v}. \ee

Obviously, system  (\ref{b18})  and operator  (\ref{b19}) can be simplified using the
 equivalence transformation (see (\ref{b35}))
 $\alpha_1v+\alpha_2 \rightarrow v$   and new notation
$\alpha_1=\frac{1}{\beta}, \ \beta\not=0 $, hence one  arrives at
the system  \be\nonumber\ba u_{xx}=d^1(u)\,u_t+C^1(u,v),
\\v_{xx}=\delta_2 v^{\beta}v_t+C^2(u,v)\ea\ee
and the operator  \be\label{b21} Q = \xi(x)\p_{x}-
 2\xi_x\frac{d^1}{d^1_u}\,\p_{u}-\frac{2\xi_x}{\beta}v\p_{v}. \ee

Substituting coefficients of   (\ref{b21})  into (\ref{b8}), we obtain  \be\label{b22}(4+\beta)\xi_{xx}=0.\ee

Solving ODE  (\ref{b22})  with $ \beta\not=-4$, we find easily
$\xi$, which must a linear function. However, the straightforward
analysis in this case   shows that  the second equation in
(\ref{b12}) is satisfied while equation  (\ref{b9})  produces
$\eta^1_{uu}=0$ (if one takes into account equations
(\ref{b13})--(\ref{b14}) and  that the function $\xi$ is linear).
 Thus, Lie symmetries can be
found only if  $ \beta\not=-4$.

Now we  examine the special case $ \beta=-4$ when  ODE (\ref{b22})
vanishes. Now  the classification equations (\ref{b9})--(\ref{b10})
(all other equations of DEs systems are  already solved) take the
form
\begin{eqnarray}&&\nonumber-
 2h\,C^1_u+\frac{1}{2}\,v\,C^1_v+2\left(1+h_u\right)C^1=-2\frac{\xi_{xxx}}{\xi_x}h+\\&&\label{b24}
\hskip2cm\frac{2}{\xi}\,\xi_{xx}(4h_u+1)h-\frac{8}{\xi^2}\,\xi^2_xh^2h_{uu},\\
&&\label{b23}
 -2h\,C^2_u+\frac{1}{2}\,v\,C^2_v+\frac{3}{2}\,
 C^2=\frac{\xi_{xxx}}{2\xi_x}\,v,
\end{eqnarray}  where  $h(u)=\frac{d^1}{d^1_u}.$
The linear first-order PDE (\ref{b23})   has the general solution  \be\nonumber
C^2=v^{-3}g(\omega)+\frac{\mu}{4}\,v,\ee
where $g$ is an arbitrary differentiable function  of the argument   $\omega=d^1v^4$. Because the function $C^2$ does not depend on $x$  the equations
\be\label{b27}\frac{\xi_{xxx}}{\xi_x}=\mu\ee
 simultaneously springs up.

In order to solve PDE (\ref{b24}) we  find its differential
consequence w.r.t. $x$:
\be\nonumber4\left(\frac{\xi^2_x}{\xi^2}\right)_xhh_{uu}-
\left(\frac{\xi_{xx}}{\xi}\right)_x(4h_u+1)=0. \ee

It turns out that  $4h_u+1\neq0 \Leftrightarrow d^1\neq\delta_1(u+\alpha_2)^{-4}$ (otherwise restrictions
 (\ref{b12})  leading to Lie symmetry  are automatically fulfilled), therefore two different cases arise:
 \[
 \hskip5mm \emph{(1)} \
\left(\frac{\xi^2_x}{\xi^2}\right)_x=0 \Rightarrow  \left(\frac{\xi_{xx}}{\xi}\right)_x=0
 \]
and \be\label{b59}\emph{(2)} \hskip3mm
\left(\frac{\xi^2_x}{\xi^2}\right)_x \not=0 \Rightarrow  4hh_{uu}-
\nu_1(4h_u+1)=0,\ee  where
$\nu_1=\left(\frac{\xi_{xx}}{\xi}\right)_x\left[\left(\frac{\xi^2_x}
{\xi^2}\right)_x\right]^{-1}.$

In  case  \emph{(1)}  we immediately obtain $\xi=\lambda_1\exp\left(\sqrt{\mu}\,x\right) $ with arbitrary
$\lambda_1$   and  $\mu>0.$ Substituting this function into PDE (\ref{b24}), the equation obtained can be solved in
 quite a similar way as PDE (\ref{b23}). Thus, we arrive at the system
\be\label{b42}\ba
u_{xx}=d^1(u)\,u_t+\frac{\lf(d^1\rg)^2}{d^1_u}f\left(v^4d^1\right)+4\mu\left(1-\frac{d^1d^1{uu}}{(d^1_u)^2}\right)\frac{d^1}{d^1_u},
\\v_{xx}=\delta_2
v^{-4}v_t+v^{-3}g\left(v^4d^1\right)+\frac{\mu}{4}\,v,\ea\ee
which is invariant under $Q$-conditional symmetry operator  \be\label{b43} Q =
\exp\left(\sqrt{\mu}\,x\right)\left(\p_{x}-
 2\sqrt{\mu}\,\frac{d^1}{d^1_u}\,\p_{u}+\frac{\sqrt{\mu}}{2}\,v\p_{v}\right), \ \mu>0. \ee

  Using the equivalence transformation (see (\ref{b35})) $x \rightarrow \frac{2}{\sqrt{\mu}}\,x, \ v \rightarrow
\sqrt[4]{\delta_2}\,v$ for system  (\ref{b42})  and operator
(\ref{b43}) we arrive at case (I) of Theorem~1.

Examination of case  \emph{(2)}  has been done in a similar way.
Taking into account (\ref{b27})  and  (\ref{b59}),  PDE  (\ref{b24})
can be again solved  \be\nonumber
C^1=\frac{\left(d^1\right)^2}{d^1_u}\left(f\left(d^1v^4\right)-\frac{\mu}{d^1}+
\left(\nu_1\frac{\xi^2_x}{\xi^2}-\frac{\xi_{xx}}{\xi}\right)\int\left(\frac{d^1_u(4h_u+
1)}{\left(d^1\right)^2}\right)du\right).\ee Because the function
$C^1$ does not depend on $x$  we obtain    ODE
\be\label{b60}\nu_1\frac{\xi^2_x}{\xi^2}-\frac{\xi_{xx}}{\xi}=\nu_2\ee
($\nu_2$  is an arbitrary constant) for the function $\xi(x)$. Thus,
one needs to solve the system of equations  (\ref{b27}) and
(\ref{b60}) in order to find the function $\xi(x)$.  If  $\nu_1=0$
then the general solution has the form  (\ref{b49*}) and then the
system and
 the operator arising in  case (II) of Theorem 1  are obtained. If  $\nu_1\not=0$ then
the general solution has the form  (\ref{b49**}) and then the system and the operator arising in  case
 (III) of Theorem 1  are obtained.

The proof is now completed. \hfill $\blacksquare$

\medskip

 Theorems 1 and 2  present   the complete (in no-go case) list of $Q$-conditional symmetries  of the first type for  the nonlinear  RD systems of the form  (\ref{b2}).
 Because each  $Q$-conditional symmetry  of the first type  is also a  $Q$-conditional symmetry
 (non-classical symmetry) the above theorems present also  a set
  (not exhaustive!) of the later symmetries for the  corresponding RD systems.
 We shall  return to this issue also in the final section.

 \section{\bf Reduction nonlinear RD system to ODE system \\ and  constructing exact solutions  }
It is well-known that using  any  $Q$-conditional symmetry
(non-classical symmetry), one reduces  the given system of PDEs  to
a system of ODEs  via  the same procedure as for classical Lie
symmetries.  Since  any   $Q$-conditional symmetry of the first type
is automatically  one of the  second type, i.e.,  the  non-classical
symmetry, we  apply  this  procedure  for finding exact solutions.
Thus, to construct an ansatz corresponding to the given operator
$Q$, the system of the linear first-order PDEs
\be\label{c1}Q(u)=0,\quad  Q(v)=0\ee
 should be solved. Substituting the ansatz obtained
into the RD system  with correctly-specified coefficients, one
obtains the reduced system of ODEs.

Let us construct  exact solutions of  the non-linear  RD system
(\ref{b30}) with $d^1=u^\beta$, \  $\beta\neq-4$ (see a comment about the case  $\beta =-4$ below). Thus, the system (\ref{b30}) takes the form
\be\label{c2}\ba u_{xx}=u^\beta
u_t+f\left(v^4u^\beta\right)u^{\beta+1}+ \frac{16u}{\beta^2},
\\v_{xx}= v^{-4}v_t+g\left(v^4u^\beta\right)v^{-3}+v,\ea\ee
Solving system (\ref{c1}) for operator (\ref{b31}), one constructs
 ansatz
\be\label{c3}\ba
u(t,x)=\varphi(t)\exp\left(-\frac{4}{\beta}\,x\right),
\\v(t,x)=\psi(t)\exp(x),\ea\ee where $\varphi(t)$ and $\psi(t)$ are
new unknown functions.

Substituting  ansatz (\ref{c3}) into (\ref{c2}), one obtains reduced
system of ODEs \be\label{c4}\ba \varphi'+f\left(\varphi^\beta
\psi^4\right)\varphi=0,\\ \psi'+g\left(\varphi^\beta
\psi^4\right)\psi=0.\ea\ee

Because the nonlinear ODEs system (\ref{c4}) is non-integrable  in
the general case, we  noted that the simplest (but still nonlinear!)
case leading to the  integrable system occurs when
 \be\label{c5}f=\alpha_1\varphi^{\beta k}
\psi^{4k}, \quad g=\alpha_2\varphi^{\beta k}
\psi^{4k},\ee where $\alpha_1, \ \alpha_2$ and $k\neq0$ are  arbitrary constants  ($k=0$
leads to  solutions of the form   $\varphi=\exp(-\alpha_1 t), \ \psi=\exp(-\alpha_2 t)$,
hence,   (\ref{c3}) will be a simple plane-wave solution of the RD system  (\ref{c2})
 with $f=\alpha_1$  and $g=\alpha_2$) .

The general solution of
(\ref{c4})  with (\ref{c5})  is  \be\label{c6}\ba
\varphi(t)=\lambda_1\left((4\alpha_2+\beta \alpha_1) k\lambda^{\beta
k}_1\,(t -t_0)\right)^{
-\frac{\alpha_1}{(4\alpha_2+\beta \alpha_1)k}},\\
\psi(t)= \left((4\alpha_2+\beta \alpha_1)k\lambda^{\beta
k}_1\,(t -t_0)\right)^{-\frac{\alpha_2}{(4\alpha_2+\beta
\alpha_1)k}},\ea\ee where $\lambda_1$ and $t_0$ are  arbitrary
constants  and  $\alpha_1\neq-\frac{4\alpha_2}{\beta}.$

The relation $\alpha_1=-\frac{4\alpha_2}{\beta}$  leads again to the plane-wave
solution of the form  mentioned above.

Thus, substituting (\ref{c6}) into ansatz (\ref{c3}) one arrives at
the two-parameter family of exact solutions
 \be\label{c9}\ba u(t,x)=\lambda_1\left((4\alpha_2+\beta \alpha_1)
k\lambda^{\beta k}_1\,(t -t_0)\right)^{
-\frac{\alpha_1}{(4\alpha_2+\beta \alpha_1)k}}\exp\left(-\frac{4}{\beta}\,x\right),\\
v(t,x)= \left((4\alpha_2+\beta \alpha_1)k\lambda^{\beta
k}_1\,(t -t_0)\right)^{-\frac{\alpha_2}{(4\alpha_2+\beta
\alpha_1)k}}\exp\left(x\right), \quad
\alpha_1\neq-\frac{4\alpha_2}{\beta}, \ k\not=0\ea\ee
of the RD system
\be\label{c8}\ba u_{xx}=u^\beta
u_t+\alpha_1u^{1+\beta(1+k)}v^{4k}+ \frac{16u}{\beta^2},
\\ v_{xx}=v^{-4}v_t+\alpha_2u^{\beta k}v^{-3+4k}+v.\ea\ee

Now we transform the RD system  (\ref{c8})  to the form, which usually
arises in applications, using the substitution
\be\label{c10} u=U^{\frac{1}{\beta+1}}, \quad v=V^{-\frac{1}{3}}\ee
by which one takes the form
 \be\label{c12}\ba
U_t=\left(U^{-\frac{\beta}{\beta+1}}U_x\right)_x-
(\beta+1)U\left(\alpha_1U^{\frac{\beta
k}{\beta+1}}V^{-\frac{4k}{3}}+
\frac{16}{\beta^2}\,U^{-\frac{\beta}{\beta+1}}\right),
\\ V_t=\left(V^{-\frac{4}{3}}V_x\right)_x+
3V\left(\alpha_2U^{\frac{\beta
k}{\beta+1}}V^{-\frac{4k}{3}}+V^{-\frac{4}{3}}\right).\ea\ee
System (\ref{c12})  can be simplified to
 \be\label{c13}\ba
U_t=\left(U^{-\kappa}U_x\right)_x-
\alpha_1^*U\left(U^{\kappa }V^{-\frac{4}{3}}\right)^k+
\frac{16}{\beta \kappa}\,U^{1-\kappa},
\\ V_t=\left(V^{-\frac{4}{3}}V_x\right)_x+ \alpha_2^* V\left(U^{\kappa }V^{-\frac{4}{3}}\right)^k
+3V^{-\frac{1}{3}}.\ea\ee
  by introducing new notations $\kappa=\frac{\beta}{\beta+1}, \  \alpha_1^*=(\beta+1)\alpha_1, \  \alpha_2^*=3\alpha_2$.
\medskip

\begin{table} \caption{Reduction of RD systems to the systems of ODEs (case \emph{(II)} of  Theorem 1) }
\begin{small}
\begin{center} \renewcommand{\arraystretch}{1}\normalsize
\begin{tabular}{|c|c|c|c| }  \hline &&&\\
  RD system&$Q$ &Ansatz & System of ODEs \\
\hline
 &&& \\ $u_{xx}=e^u u_t+e^uf\left(\omega\right)$ &$\left(e^{2x}+\alpha
e^{-2x}\right)\p_x-$&
 $u=\varphi(t)-$&$e^\varphi\varphi'+e^\varphi f(\chi)+32\alpha=0$\\
 $v_{xx}=v^{-4}v_t+v+$&$4\left(e^{2x}-\alpha e^{-2x}\right)\p_u+
$&$2\ln \left(e^{2x}+\alpha e^{-2x}\right) $&$\psi'+\psi
g(\chi)-4\alpha\psi^5=0$\\$v^{-3}g\left(\omega\right), \
\omega=e^uv^4$&$\left(e^{2x}-\alpha
e^{-2x}\right)v\p_v$&$v=\psi(t)\left(e^{2x}+\alpha
e^{-2x}\right)^{\frac{1}{2}}$&$\chi=e^{\varphi}\psi^4$\\
  \hline
  &&& \\ $u_{xx}=e^u u_t+e^uf\left(\omega\right)$ &$\sin (2x)\p_x-$&
 $u=\varphi(t)-2\ln \sin (2x)$&$e^\varphi\varphi'+e^\varphi f(\chi)-8=0$\\
 $v_{xx}=v^{-4}v_t-v+$&$4\cos (2x)\p_u+
$&$v=\psi(t)\sin^{\frac{1}{2}}(2x)
 $&$\psi'+\psi g(\chi)+\psi^5=0$\\$v^{-3}g\left(\omega\right), \ \omega=e^uv^4$&$\cos (2x)\,v\p_v$&&
 $\chi=e^{\varphi}\psi^4$\\
  \hline

  &&& \\ $u_{xx}=u^\beta
u_t+$ &$\left(e^{2x}+\alpha e^{-2x}\right)\p_x-$&
 $u=\varphi(t)\left(e^{2x}+\alpha e^{-2x}\right)^{-\frac{2}{\beta}}$&$\varphi'+\varphi f(\chi)+$\\
 $u^{\beta+1}f\left(\omega\right)+ \frac{16}{\beta^2}u
$&$\frac{4}{\beta}\left(e^{2x}-\alpha e^{-2x}\right)u\p_u
 $&$v=\psi(t)\left(e^{2x}+\alpha e^{-2x}\right)^{\frac{1}{2}}$&$\frac{32\alpha(2+\beta)}{\beta^2}\varphi^{1-
 \beta}=0$\\
 $v_{xx}= v^{-4}v_t+v+$&$\left(e^{2x}-\alpha e^{-2x}\right)v\p_v$&&$\psi'+\psi g(\chi)-4\alpha\psi^5=0$\\
 $v^{-3}g\left(\omega\right), \ \omega=u^\beta v^4$&&&$\chi=\varphi^{\beta}\psi^4$\\
  \hline

  &&& \\ $u_{xx}=u^\beta
u_t+$ &$\sin (2x)\p_x-$&
 $u=\varphi(t)\sin^{-\frac{2}{\beta}}(2x)$&$\varphi'+\varphi f(\chi)-$\\
 $u^{\beta+1}f\left(\omega\right)- \frac{16}{\beta^2}u
$&$\frac{4}{\beta}\cos (2x)u\p_u+
 $&$v=\psi(t)\sin^{\frac{1}{2}}(2x)$&$\frac{8(2+\beta)}{\beta^2}\varphi^{1-
 \beta}=0$\\
 $v_{xx}= v^{-4}v_t-v+$&$\cos (2x)v\p_v$&&$\psi'+\psi g(\chi)+\psi^5=0$\\
 $v^{-3}g\left(\omega\right), \ \omega=u^\beta v^4$&&&$\chi=\varphi^{\beta}\psi^4$\\
  \hline
   \end{tabular}
\end{center}
\end{small}
\end{table}

  \textbf{Remark 2.} The RD system  (\ref{c13})  with
  $\kappa=\frac{4}{3}$,
  is a particular case of one (28) \cite{ch-king4}, hence,  the terms containing $U^{-\frac{1}{3}}$
  and $V^{-\frac{1}{3}}$ can be removed by the substitution (34)\cite{ch-king4}.
  However, this substitution does not work for $\kappa \not=\frac{4}{3}$.

  Now we rewrite (\ref{c9}) using substitution (\ref{c10}) and new
   notations, hence, the two-parameter family of exact solutions takes the form
   \be\label{c14}\ba U(t,x)=\lambda_1\left(\gamma
k\lambda^{\beta k}_1\,(t -t_0)\right)^{
-\frac{\alpha_1^*}{\gamma k}}\exp\left(-\frac{4}{\kappa}\,x\right),\\
V(t,x)= \left(\gamma k\lambda^{\beta
k}_1\,(t -t_0)\right)^{-\frac{\alpha_2^*}{\gamma k}}\exp(-3x),
\ea\ee
where $\gamma k\not=0, \ \gamma=\kappa \alpha_1^*+ \frac{4}{3} \alpha_2^*. $

It should be noted that  solutions  of the form  (\ref{c14})  possess essentially different properties
depending on the parameter signs, namely:
 \begin{enumerate}
            \item  each  solution $(U,V)$   blows up for the finite time $t_0>0$
            provided $\alpha_1^* \gamma k>0$  and $\alpha_2^* \gamma
            k>0$;

                 \item the component $U$ blows up for the finite time $t_0>0$ while the component $V$
                  vanishes provided $\alpha_1^* \gamma k>0$  and $\alpha_2^* \gamma k<0$ (and vice
                  versa);
                     \item  each  solution $(U,V)$ vanishes for the finite time $t_0>0$
                      provided $\alpha_1^* \gamma k<0$  and $\alpha_2^* \gamma
                      k<0$;
                         \item  if $t_0<0$ then all solutions belonging to the family
                         (\ref{c14}) are global (in time) and, depending on the parameters and the space
                          domain, can be either bounded or unbounded.
  \end{enumerate}

 Thus,   the  solutions obtained
 may be used for  describing  a wide range of  different processes  arising in applications (examples  lie outside the scope of the current work).
However,  it should be stressed that each solution of the form  (\ref{c14})  is a non-Lie solution,
  i.e. one is not obtainable via Lie symmetry operators. In fact, the nonlinear RD system
   (\ref{c13}) under the restrictions listed above admits only the trivial Lie algebra
   generated by time  and space  translation operators (see Theorem 1 in \cite{ch-king4}).
   It means that plane-wave solutions only are obtainable via the Lie method.

 Exact solutions of  non-linear  RD systems arising in cases (II) and (III) of Theorem~1
 can be constructed in a similar way.
In particular, using operator   (\ref{b33*}),    four different
reductions  were obtained for the RD systems of the form
(\ref{b32*}), which are listed in Table 1 (the parameter $\mu$ can
be reduced to $\pm 4 $ without losing a generality). One observes
that structures of solutions essentially depend on diffusivity in
the first equation and on   the linear term sign in the second
equation.  Because the ODE systems obtained have rather complicated
structures, one needs to specify correctly  the functions $f$ and
$g$ in order to find exact solutions in an explicit  form (otherwise
numerical simulations should be applied).

 \section{\bf Discussion }

In this paper, $Q$-conditional symmetries  of the form (\ref{b1})
for the RD systems belonging to the class  (\ref{b2}) and their
application are studied. The main result is presented in Theorems 1
and 2. Thus, taking into account Theorem 1 \cite{che-dav2012}, one
may claim that $Q$-conditional symmetries of the first type of the
RD systems with $(d^1_u)^2+(d^2_v)^2\neq0$ are completely described.
In particular, three new subclasses of RD systems  with non-constant
diffusivities are found (see Theorem 1), which do not coincide with
those presented in \cite{che-dav2012} (see Table 1). It is interesting to
note that all the RD systems listed in Theorem 1 contain the second
RD equation involving the power-law diffusivity with exponent $-4$.
This exponent corresponds to the known critical exponent
$-\frac{4}{3}$ arising in RD systems  of the form (\ref{1}) with the
widest  Lie symmetry  \cite{ch-king4,ibrag-94}(note that the list of the RD systems possessing non-trivial Lie symmetry was essentially reduced in \cite{ch-king4} comparing with earlier paper \cite{ibrag-94}
 in order to obtain really in-equivalent systems). Moreover, setting
formally the diffusivity with the same exponent (i.e. $d^1=u^{-4}$)
in the first equation, one arrives at the known (from the Lie
symmetry point of view!) systems. Thus, the systems obtained here
are generalizations of those with the widest Lie symmetry in order
to obtain conditional symmetry instead of classical symmetry.

It is well-known that the differential consequences of the operator
in question can be  used  in order to extend non-classical symmetry
(see, e.g., \cite{zh-lahno98,fss}), hence
 one may reformulate Definition~2 in a such way that
 the manifold
 \[{\cal{M}}^*_2=\{S_1=0, S_2=0, Q(u)=0, Q(v)=0, \frac{\partial}{\partial t}Q(u)=0, \frac{\partial}{\partial x}Q(u)=0,
  \frac{\partial}{\partial t}Q(v)=0,
  \frac{\partial}{\partial x}Q(v)=0 \}
  \]
   can
  be used instead of ${\cal{M}}_2 $.
   However, the definition
   obtained does not lead to any new conditional
    symmetries of system (\ref{b2}) if one is
    looking for  operators of the form  (\ref{b3})
    with $\xi^0\neq0$ \cite{ch-2010} (the detailed
     proof is presented in \cite{pliukhin}).
     In the case $\xi^0=0, \xi^1\neq0$ the situation
     is essentially different. As it is shown above,
     Definition 2 produces   the system of determining
     equations (\ref{b57}) (this is still  a   challenge  to solve one without any restrictions).
 On the other hand,  application
 of this definition with the manifold
 ${\cal{M}}^*_2 $ leads to a system of determining
 equations, which  consists of two equations only,
 namely (in this case we can put $\xi^1=1$ without losing a generality):
\begin{eqnarray}
&&\nonumber\eta^1C^1_u+\eta^2C^1_v+
\eta^1\frac{d^1_u}{d^1}\left(\eta^1_x+\eta^1\eta^1_u+
\eta^2\eta^1_v-C^1\right) -\frac{d^1}{d^2}\eta^1_vC^2-\eta^1_uC^1 +
d^1\eta^1_t-\eta^1_{xx}+ \\
&&\nonumber\hskip2cm\left(\frac{d^1}{d^2}-
1\right)\eta^1_v\left(\eta^2_x+\eta^1\eta^2_u+\eta^2\eta^2_v\right)-
2\eta^1\eta^1_{xu}-2\eta^2\eta^1_{xv}-(\eta^1)^2\eta^1_{uu}-\\&&
\label{d1}
\hskip5cm 2\eta^1\eta^2\eta^1_{uv}-(\eta^2)^2\eta^1_{vv}=0,\\
&&\nonumber \eta^1C^2_u+\eta^2C^2_v+
\eta^2\frac{d^2_v}{d^2}\left(\eta^2_x+\eta^1\eta^2_u+
\eta^2\eta^2_v-C^2\right) -\frac{d^2}{d^1}\eta^2_uC^1-\eta^2_vC^2 +
d^2\eta^2_t-\eta^2_{xx}+ \\
&&\nonumber\hskip2cm\left(\frac{d^2}{d^1}-
1\right)\eta^2_u\left(\eta^1_x+\eta^1\eta^1_u+\eta^2\eta^1_v\right)-
2\eta^1\eta^2_{xu}-2\eta^2\eta^2_{xv}-(\eta^1)^2\eta^2_{uu}-\\&&
\nonumber \hskip5cm
2\eta^1\eta^2\eta^2_{uv}-(\eta^2)^2\eta^2_{vv}=0.
\end{eqnarray}
According to the so called no-go theorem \cite{zh-lahno98} (its generalisation on  systems of evolution equations is straightforward), system (\ref{d1}) is reducible  to the initial RD system (\ref{b3}), i.e. cannot be solved in the general case.
  In other words, the amended definition is not applicable for constructing new $Q$-conditional symmetries because the problem is reduced to solving the  reaction-diffusion system in question.
However,  we have shown in this paper  that using Definition~1 a special subset of such symmetries can be completely
   described and found in explicit form.

   Generally  speaking, the notion  of  $Q$-conditional symmetry of the $p$-th  type \cite{ch-2010}, which  is  successfully applied here to the nonlinear RD systems  of  the form  (\ref{b2}),
   may be thought as a further development of the concept of conditional invariance proposed
    in \cite [Section 5.7]{fss}
    (see also highly nontrivial examples in \cite{ch-he-2004}). It is important because a list of  successful applications of this concept  for nonlinear systems of evolution equations  is relatively short.

    It should be also noted that symmetry based methods for solving nonlinear PDEs have clear connection to the method of differential
constraints, which has been formulated in \cite{yanenko64} (see also the  later monograph \cite{yanenko84}). In fact,
     a common property that underlies  all the   symmetry based methods  can be described as follows: in order to find  exact  solutions one solves   a  nonlinear PDE (system of PDEs)  together with the differential
constraint(s) generated by a symmetry operator. The corresponding
symmetry can be of different types: Lie symmetry, $Q$-conditional
symmetry, generalised  conditional symmetry etc. Because the
over-determined system  consisting of  the given PDE and the
differential constraint  is compatible one  can find  its solutions
in a much simpler way. It means that  the main problem of the method
of differential constraints, how to define suitable constraint(s)
for the given PDE  in a such way that the over-determined system
obtained will be compatible,  is automatically solved. Of course,
one may try to find  the suitable differential  constraints by other
methods (see, e.g., \cite{ch98-c,c-fusco-m-03} and the papers cited
therein), i.e. without using any symmetry based approach. An
overview of possible approaches with attempt to create the general
algorithm of integrating over-determined systems is presented in
\cite{yanenko84}.

In order to demonstrate the applicability of the derived symmetries,
we used those for reducing
 the nonlinear RD systems to the relevant ODEs systems and constructing exact solutions.
 In particular   $Q$-conditional operators arising in  cases (I) and (II) of Theorem~1 were
 used in order  to construct
non-Lie ans\"atze and to reduce the relevant RD systems  to  the
corresponding ODE systems, which are presented in formulae
(\ref{c3})--(\ref{c4}) and  Table~1. As result,
 multi-parameter  families  of exact solutions in  the explicit  form  (\ref {c14})   were  constructed
 for the RD system (\ref {c13})  with an arbitrary power-law  diffusivity. Moreover, we
 have shown that the  solutions obtained  possess attractive properties, hence
 can  describe different phenomena arising in applications.

\section{Acknowledgements}
This research was partly supported  to the first author  by a Marie Curie International Incoming Fellowship within the 7th European
Community Framework Programme.

\end{document}